\documentclass[twocolumn]{aastex631}
\usepackage{amsmath,amssymb,bm,latexsym}

\usepackage{xcolor}
\usepackage[normalem]{ulem} 
\begin{document}

\newcommand*{\mkblue}[1]{{\color{blue}{#1}}}
\title{Challenges in Binary Pulsar Timing Detection of Dark Matter Subhalos}

\author[0009-0001-9956-6375]{Zheng-Long Wang}
\affiliation{Key Laboratory of Dark Matter and Space Astronomy, Purple Mountain Observatory, Chinese Academy of Sciences, Nanjing 210033, People's Republic of China}
\affiliation{School of Astronomy and Space Science, University of Science and Technology of China, Hefei, Anhui 230026, People's Republic of China}

\author[0000-0003-4963-7275]{Zi-Qing Xia}
\email{xiazq@pmo.ac.cn}
\affiliation{Key Laboratory of Dark Matter and Space Astronomy, Purple Mountain Observatory, Chinese Academy of Sciences, Nanjing 210033, People's Republic of China}

\author[0000-0002-7275-8561]{Yue-Lin Sming Tsai}
\email{smingtsai@pmo.ac.cn}
\affiliation{Key Laboratory of Dark Matter and Space Astronomy, Purple Mountain Observatory, Chinese Academy of Sciences, Nanjing 210033, People's Republic of China}
\affiliation{School of Astronomy and Space Science, University of Science and Technology of China, Hefei, Anhui 230026, People's Republic of China}

\author[0000-0002-8966-6911]{Yi-Zhong Fan}
\email{yzfan@pmo.ac.cn}
\affiliation{Key Laboratory of Dark Matter and Space Astronomy, Purple Mountain Observatory, Chinese Academy of Sciences, Nanjing 210033, People's Republic of China}
\affiliation{School of Astronomy and Space Science, University of Science and Technology of China, Hefei, Anhui 230026, People's Republic of China}

\begin{abstract}
Recently, binary pulsar timing has been proposed as a viable probe of dark matter subhalos with masses of $\sim 10^7\,M_{\odot}$ in the solar neighborhood. 
First, we present a comprehensive analytical framework that incorporates the subhalo mass function, projection effects of LOS acceleration, and the spatiotemporal geometric requirements for joint detection by binary systems, enabling a quantitative evaluation of the detectability of nearby subhalos. 
Applying this framework to the current binary pulsar sample, 
we find 
a probability $\leq 1.7 \times 10^{-4}$ of detecting at least one subhalo within the effective volume.
Second, an independent timing residual analysis shows no statistically significant excess in LOS accelerations beyond predictions from data-driven Galactic gravitational potential models. 
These results place stringent constraints on detecting $<10^8~M_{\odot}$ dark matter subhalos with existing pulsar timing data, 
aligning with the theoretical expectation that such subhalos have a low survival probability in the solar neighborhood.
A low detection prospect still holds even for future Square Kilometre Array observations.
\end{abstract}

\section{Introduction}
In the standard cosmological model, structure formation proceeds via a hierarchical ``bottom-up'' assembly process~\citep{1978MNRAS.183..341W,zavala2019darkmatterhaloessubhaloes}. As a direct consequence, dark matter (DM) halos of Milky Way mass are predicted to host a plethora of substructures, known as subhalos~\citep{Springel_2008,Barry_2023}. 
Most of these low-mass subhalos lack stellar components, 
making them inaccessible to direct electromagnetic observations and leaving them as distinct “dark” structures~\citep{2000ApJ...539..517B,Sawala_2017,Fitts_2017}.
These invisible substructures encode critical information regarding small-scale primordial density fluctuations and serve as vital probes for distinguishing cold DM 
from alternative scenarios~\citep{Sawala_2017,2025ApJ...978L..23Z,Lei_2025}. Therefore, direct detection of low-mass DM subhalos via gravitational effects (e.g., stellar streams~\citep{Erkal_2017,Bonaca_2019,2025ApJ...978L..23Z} and strong gravitational lensing~\citep{Vegetti_2012,Sawala_2017,Lei_2025}) is of fundamental importance for validating theories of structure formation and constraining the DM particle nature.

Recently, binary pulsar timing has been proposed to detect local DM subhalos with masses of $\sim 10^7\,M_\odot$~\citep{2026PhRvL.136d1201C}. 
The precisely measured time derivative of the orbital period provides a robust tracer of gravitational acceleration and 
has been widely used for precision accelerometry~\citep{Phillips_2021,Chakrabarti_2021,Moran_2024,2024PhRvD.110b3026D}.
Assuming no mass transfer in the binary system, the time derivative of the observed orbital period can be decomposed as~\citep{2024PhRvD.110b3026D},
\begin{equation}
    \dot{P}_{\rm acc} = \dot{P}_{\rm obs} - \dot{P}_{\rm GW} - \dot{P}_{\rm shk} ,
    \label{eq1}
\end{equation}
where $\dot{P}_{\rm obs}$ is the observed rate.
The term $\dot{P}_{\rm GW}$ represents the orbital decay driven by gravitational-wave emission, 
whereas $\dot{P}_{\rm shk}$ denotes the apparent Doppler contribution induced by transverse motion (the Shklovskii effect~\citep{Shklovskii1970}).
The most critical contribution, $\dot{P}_{\rm acc}$, corresponds to the line-of-sight (LOS) acceleration, 
arising from the Galactic background potential, possible DM subhalos, or other external gravitational sources. 
Subtracting the Galactic contribution allows, in principle, the extraction of any residual LOS acceleration induced by unknown mass distributions. However, an apparent acceleration anomaly in a single binary system is generally indistinguishable from statistical fluctuations. 
In contrast, a spatially correlated signal observed in two nearby binaries may provide compelling evidence for a common gravitational perturber, such as a subhalo.

A candidate DM subhalo with a mass of $6\times 10^{7}~M_\odot$ has been reported at a distance of $\sim 0.78$ kpc from the solar system~\citep{2026PhRvL.136d1201C}. If confirmed, such an object would have significant implications for physics and astrophysics, ranging from the DM particle nature to galaxy formation. Its inferred DM annihilation $J$-factor could reach $\sim 10^{23}~{\rm GeV^{2}\,cm^{-5}}$, comparable to that of the Galactic Center~\citep{2025arXiv251213261Z}. Dynamically, a subhalo of this mass might also generate phase-space spirals within the Galactic disk~\citep{2026arXiv260309015F}.

In this Letter, combining simulations with observations, we demonstrate that capturing subhalo signals with binary pulsars in the solar neighborhood is very challenging under reasonable cold DM model assumptions. 
As a verification, we re-analyze the binary pulsar data using the approach of \cite{2026PhRvL.136d1201C}, but with proper treatment of pulsar distance errors and data-driven Galactic gravitational potential models.
No credible signal of subhalos was identified in this specific analysis, in agreement with our general analysis.

\begin{figure*}[t!]
    \centering
    \includegraphics[width=\textwidth]{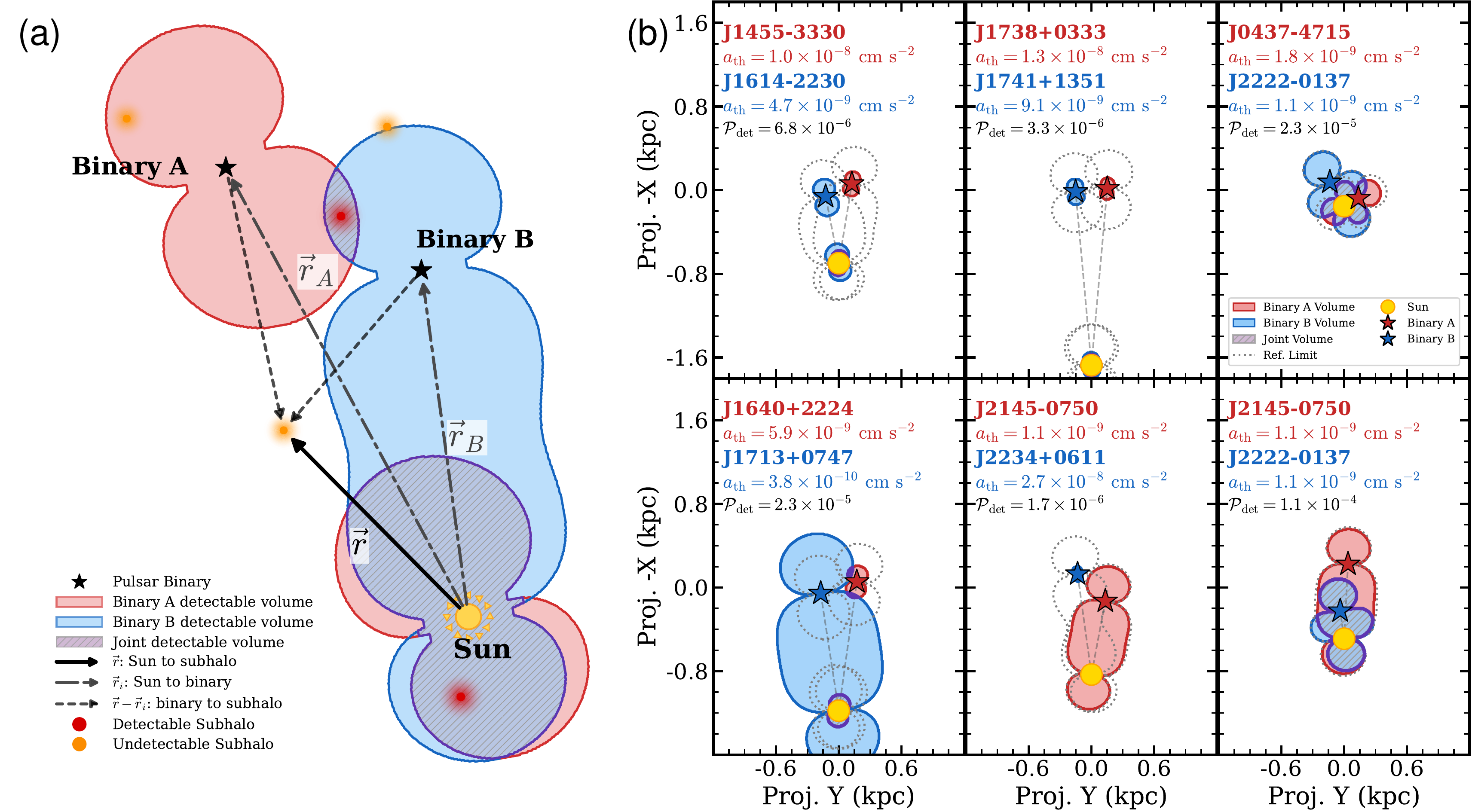}
\caption{
Geometric configuration and effective detection volumes for joint subhalo detection by binary pulsars.
\textbf{Left panel:} Detection method. The red and blue shaded regions represent the detectable volumes of Binary A and Binary B, respectively, while the purple shading shows their overlapping joint detectable volume.
Red dots indicate detectable subhalos, while orange dots indicate undetectable subhalos.
\textbf{Right panels:} Detection geometry for the six neighboring binary pulsar pairs. Red and blue regions show the sensitivity lobes of each pulsar to a $10^7\,M_\odot$ subhalo under current precision, and purple regions mark the joint detectable volumes. Yellow dots denote the Sun, and red/blue stars denote the two pulsars. Gray dotted contours indicate the reference detection regions for an ideal acceleration sensitivity of $10^{-9}\,{\rm cm\,s^{-2}}$. The acceleration thresholds and corresponding joint detection probabilities are labeled in each panel.
}

    \label{fig:subhalo_joint_detection}
\end{figure*}

\section{Subhalo Detection Probability}

We quantify the detectability of subhalos with binary pulsars, 
incorporating the subhalo mass function, LOS acceleration projection effects, and spatiotemporal geometric requirements for joint detection. 
Based on DM-only simulations, we assume a power-law subhalo mass distribution in the Galactic halo~\citep{Springel_2008}:
\begin{equation}
\frac{dn}{dM} = A \left( \frac{M}{M_{\odot}} \right)^{-1.9},
\end{equation}
where $n$ is the number density. 
The normalization $A$ is set to the maximum allowed by the \texttt{FIRE-2} and \texttt{APOSTLE} simulations 
within 3~kpc of the Sun (the region hosting high-precision binary pulsars), the number density of subhalos with $M > 10^{6.5}\,M_\odot$ is at most $10^{-3}\,\mathrm{kpc}^{-3}$~\citep{Barry_2023,Sawala_2017}. 
This parameterization neglects baryonic effects (e.g., tidal stripping, and disk shocking~\citep{Garrison_Kimmel_2017,Kelley_2019}) and small-scale suppression from alternative DM models (e.g., warm, self-interacting, and fuzzy DM~\citep{Lovell_2014,Tulin_2018,Du_2016}), 
which may lower the subhalo abundance well below this bound throughout the Milky Way. 
Hence, our detection probability represents an optimistic upper limit, potentially overestimating the true rate by up to an order of magnitude.


Binary pulsar timing effectively measures the differential LOS acceleration between the pulsar and the Solar System barycenter~\citep{2026PhRvD.113f3033D}. Adopting the heliocentric coordinate system with the Sun at the origin, let $\mathbf r$ denote the subhalo position relative to the Sun, and $\mathbf r_i$ the position of the $i$-th pulsar relative to the Sun. For a subhalo of mass $M$, the induced differential LOS acceleration is
\begin{equation}
a_{{\rm LOS},i}(\mathbf r,M) =
GM \left[
\frac{\mathbf r-\mathbf r_i}{|\mathbf r-\mathbf r_i|^3} - \frac{\mathbf r}{|\mathbf r|^3}
\right] \cdot \hat{\mathbf n}_i,
\qquad
\hat{\mathbf n}_i = \frac{\mathbf r_i}{|\mathbf r_i|},
\label{eq:subhalo_diff_accel}
\end{equation}
where the first term represents the acceleration of the pulsar due to the subhalo, and the second term represents the acceleration of the Sun.

Given the acceleration threshold $a_{{\rm th},i}$ of the $i$-th pulsar,  the set of detectable spatial positions for a subhalo of mass $M$ is defined as

\begin{equation}
\mathcal V_i(M,a_{{\rm th},i}) =
\left\{ \mathbf r \in \mathbb{R}^3 \,\middle|\,
|a_{{\rm LOS},i}(\mathbf r,M)| \ge a_{{\rm th},i}
\right\}.,
\label{eq:single_volume}
\end{equation}
as illustrated in Fig.~\ref{fig:subhalo_joint_detection}. The sensitivity region of a single pulsar forms two axisymmetric lobes near the pulsar and near the Sun. When pulsars are sufficiently close and acceleration thresholds are low, these lobes may connect. Only subhalos located in the overlapping region of two pulsars can produce a joint detection signal.

In practice, possible non-subhalo gravitational perturbations may contaminate individual pulsar timing signals. As illustrated in Fig.~\ref{fig:subhalo_joint_detection}, a robust detection requires joint observations of two nearby binary pulsars (A and B), using spatial correlations to suppress noise and break positional degeneracies~\citep{2026PhRvL.136d1201C}. The effective volume for joint detection is the geometric overlap of their individual detection regions,
\begin{equation}
V_{\rm eff}(M) = V[\mathcal{V}_A(M, a_{{\rm th},A}) \cap \mathcal{V}_B(M, a_{{\rm th},B})].
\label{eq:Veff}
\end{equation}

The expected number of detectable subhalos for a binary pair follows from integrating over the mass function,
\begin{equation}
N_{\rm exp} = \int_{0}^{M_{\rm max}} \frac{dn}{dM} \, V_{\rm eff}(M) \, dM,
\label{eq:Nexp}
\end{equation}
where we set $M_{\rm max}=10^8\,M_\odot$, as subhalos above this mass are expected to host stars~\citep{Barry_2023} and are extremely rare (fewer than 0.1 within 10 kpc of the Galactic center~\citep{Barry_2023,Sawala_2017}). 
Assuming the number of detections follows Poisson statistics and summing over all possible nonzero detection, 
the probability of detecting at least one subhalo is
\begin{equation}
\mathcal{P}_{\rm det} = 1 - \exp(-N_{\rm exp}),
\end{equation}
where the exponential term corresponds to the probability of no detection.

We adopt the latest binary pulsar dataset~\citep{2025PhRvD.111j3036D}, excluding spider pulsars, globular cluster members, and systems with large measurement errors (J1603$-$7202, J1017$-$7156~\citep{Reardon_2021}, and J1012$-$4235~\citep{2024A&A...682A.103G}), resulting in 24 high-precision binary systems. To account for distance uncertainties, we perform Monte Carlo (MC) sampling of observed parallaxes $\varpi$ and reconstruct pulsar distances via $d=1/\varpi$, determining relative positions between binaries. The acceleration threshold for each pulsar is set by the fractional uncertainty in its orbital period derivative
\begin{equation}
a_{\rm th} = c\,\sigma_{\rm acc},
\end{equation}
where $c$ is the speed of light, $P_b$ is the orbital period, and $\sigma_{\rm acc}$ is the uncertainty of $\dot P_{\rm acc}/P_b$, with $\dot P_{\rm acc}$ given by Eq.~\eqref{eq1}. This uncertainty is propagated from $\dot P_{\rm obs}$, $\dot P_{\rm GW}$, $\dot P_{\rm shk}$, and $P_b$. Using a $1\sigma$ threshold, this remains an optimistic detection criterion.


For a joint detection, the detectable regions of two binary pulsars must overlap. 
The Sun-proximate lobes of two pairs always overlap.
If a subhalo is very close to the Sun, it induces differential acceleration signals in all directions, not just along a specific binary. Examining the pulsars with the lowest acceleration thresholds (J1713$+$0747, J2222$-$0137, and J0437$-$4715), none show acceleration anomalies. All other pulsars, except for the similarly sensitive J2145$-$0750, have higher thresholds and correspondingly smaller detectable volumes, which are fully covered by the three high-precision binaries without anomalies. Based on this, we do not extend candidate pair selection using Sun-proximate lobes but retain the 0.5~kpc spacing criterion from~\cite{2026PhRvL.136d1201C} to select spatially neighboring binaries that may share a local perturbation on one side. Pairs separated beyond this distance showing acceleration anomalies are more likely affected by independent perturbations rather than a common subhalo. This criterion identifies six neighboring binary pairs (see Table~\ref{tab:residuals_sigma}). We compute $V_{\rm eff}(M)$ using Eq.~\eqref{eq:subhalo_diff_accel} and the actual spatial configuration of each pair.

As shown in Fig.~\ref{fig:subhalo_joint_detection}, accounting for realistic geometry and individual pulsar sensitivities drastically reduces the joint detection volume. To avoid double-counting volumes from different pairs, we use the union of all pair volumes to compute the cumulative detection probability for the full sample, yielding
\begin{equation}
\mathcal{P}_{\rm det}  \lesssim 1.7 \times 10^{-4}.
\label{eq:plimit}
\end{equation}
Therefore, we conclude that detecting dark matter subhalos in the solar neighborhood with current binary pulsar timing is theoretically highly improbable.

\begin{table*}[t!]
\centering

\caption{Parameters for the six neighboring binary pulsar pairs with separation less than $0.5~\text{kpc}$. 
Distances (pair separations) and uncertainties are derived from the 16th/84th percentiles of the MC sampling distribution. 
For each pair, the columns list the significance $\sigma$ of the deviation from three different Galactic gravitational potential models 
(\textbf{MW2014}~\citep{Bovy_2015}, \textbf{Mc2017}~\citep{McMillan_2016}, and \textbf{Gala2022})~\citep{PriceWhelan2022} for the first (P1) and second (P2) pulsar.}
\label{tab:residuals_sigma}
\begin{tabular}{lcccc}
\hline
\textbf{Pulsar Pair (P1 \& P2)} & \textbf{Separation (kpc)} & \textbf{MW2014} ($\sigma_{P1} / \sigma_{P2}$) & \textbf{Mc2017} ($\sigma_{P1} / \sigma_{P2}$) & \textbf{Gala2022} ($\sigma_{P1} / \sigma_{P2}$) \\
\hline
J1455$-$3330 \& J1614$-$2230 & $0.28_{-0.03}^{+0.04}$ & 1.78 / 0.66 & 1.76 / 0.69 & 1.80 / 0.75 \\
J0437$-$4715 \& J2222$-$0137 & $0.31_{-0.00}^{+0.00}$ & 0.55 / 0.29 & 0.52 / 0.14 & 0.68 / 0.47 \\
J2145$-$0750 \& J2234$+$0611 & $0.37_{-0.03}^{+0.04}$ & 1.80 / 0.47 & 1.46 / 0.48 & 1.67 / 0.45 \\
J1738$+$0333 \& J1741$+$1351 & $0.38_{-0.07}^{+0.18}$ & 0.19 / 1.53 & 0.23 / 1.54 & 0.22 / 1.55 \\
J1640$+$2224 \& J1713$+$0747 & $0.38_{-0.05}^{+0.28}$ & 2.04 / 0.70 & 2.06 / 0.47 & 1.93 / 0.29 \\
J2145$-$0750 \& J2222$-$0137 & $0.45_{-0.04}^{+0.04}$ & 1.80 / 0.29 & 1.46 / 0.14 & 1.67 / 0.47 \\
\hline
\end{tabular}
\end{table*}

=\section{Timing Residual analysis}
Motivated by the stringent theoretical limit in Eq.~\eqref{eq:plimit}, we reanalyze the observational data~\citep{2025PhRvD.111j3036D} by computing the residual orbital period derivative $\dot{P}{\rm res}$ for the six neighboring binary pulsar pairs in Table~\ref{tab:residuals_sigma}. 
This residual isolates the contribution of unseen perturbers can be written as
\begin{equation}
    \dot{P}_{\rm res} = (\dot{P}_{\rm obs} - \dot{P}_{\rm GW} - \dot{P}_{\rm shk}) - \dot{P}_{\rm Gal}.
    \label{eq:pres}
\end{equation}
Under a smooth Galactic model and in the absence of external perturbers, 
$\dot{P}_{\rm res}$ is expected to be zero.

We follow ~\cite{2026PhRvL.136d1201C} to evaluate each term in Eq.~\eqref{eq:pres}.
\begin{itemize}
    \item $\dot{P}_{\rm obs}$ and $\dot{P}_{\rm GW}$: Central values and measurement uncertainties are taken from the high-precision pulsar dataset~\citep{2025PhRvD.111j3036D}. In the MC analysis, they are sampled from a Gaussian distribution.

    \item $\dot{P}_{\rm shk}$: The term accounts for the Shklovskii effect—an apparent acceleration induced by the pulsar's transverse motion~\citep{2026PhRvL.136d1201C}:
    \begin{equation}
        \dot{P}_{\rm shk} = \frac{P_b \mu^2 d}{c},
    \end{equation}
    where $\mu$ is the total proper motion.
    
    \item $\dot{P}_{\rm Gal}$: This term represents the Galactic potential correction. 
    We cross-validate it using three standard potential models: \textbf{MW2014}~\citep{Bovy_2015}, \textbf{Gala2022}~\citep{PriceWhelan2022}, and \textbf{Mc2017}~\citep{McMillan_2016}. 
    Particularly, the \textbf{Mc2017} model uniquely accounts for potential uncertainties through its sampling. 
    For the Galactic potential, we adopt the solar position in the Galactocentric coordinate system 
    as $(x_\odot, y_\odot, z_\odot) = (-8.178,\, 0.0,\, 0.0208)$~kpc~\citep{Gravity2019_R0, Bennett2019_SunHeight}.
\end{itemize}

Conventional methods typically rely on simple Gaussian Error Propagation (GEP) based on parallax measurements~\citep{2024PhRvD.110b3026D}. However, due to the non-linear relationship $d = 1/\varpi$, a symmetric Gaussian error in $\varpi$ results in a distance probability distribution with an asymmetric ``long-tail'' for sources with low parallax precision. 
To rigorously address these uncertainties, we implement a unified MC framework that simultaneously samples all observables.

The accuracy of $\dot{P}_{\rm res}$ is strongly governed by the precision of the distance $d$, 
which simultaneously determines both the Shklovskii contribution and the Galactic potential correction. 
To properly account for this, in each of the $\mathcal{N}_{\text{total}} = 200,000$ realizations, 
we independently sample $\dot{P}_{\rm obs}$ and $\dot{P}_{\rm GW}$. 
In each iteration, we also sample the parallax $\varpi$ to derive the distance $d$, 
which is then used to compute $\dot{P}_{\rm shk}$ and $\dot{P}_{\rm Gal}$.

Because the resulting distribution of $\dot{P}_{\rm res}$ inherits significant asymmetry from the distance inversion, 
we employ a shape-independent quantile counting method to define the Gaussian-equivalent significance in units of $\sigma$
\begin{equation}
    \mathcal{P}(x > 0) = \frac{\mathcal{N}(x_i > 0)}{\mathcal{N}_{\text{total}}},
\end{equation}
where $x$ represents the residual value and $x_i$ denotes its value in the $i$-th MC realization. The two-sided significance is then derived as
\begin{equation}
    \mathcal{P}_{\text{two-sided}} = 2 \cdot \min\left[\mathcal{P}(x>0),\, 1-\mathcal{P}(x>0)\right],
\end{equation}
where its significance is $\sqrt{2} \operatorname{erf}^{-1}(1 - \mathcal{P}_{\text{two-sided}})$.
This statistical definition ensures that our assessment remains robust and conservative, 
even for non-Gaussian or highly asymmetric distance errors.

\begin{figure}[t!]
    \centering
    \includegraphics[width=0.95\columnwidth]{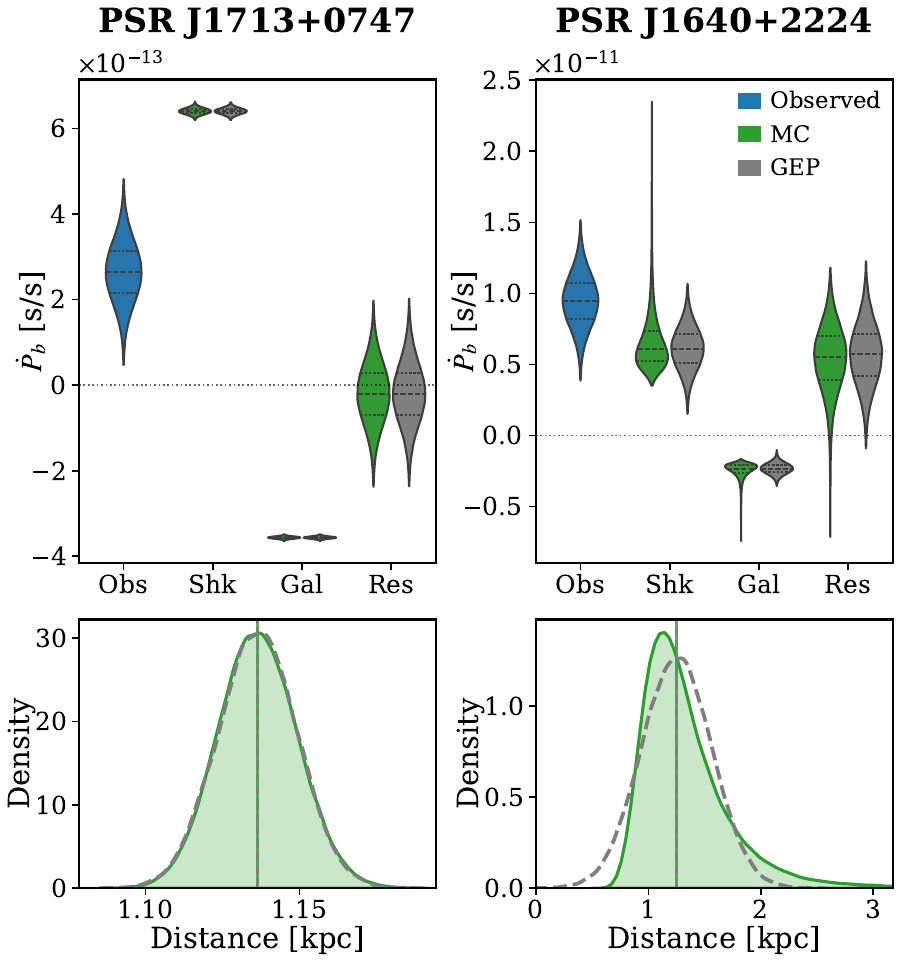} 
\caption{Comparison of distance and residual distributions for high-precision (\textbf{J1713+0747}) and low-precision (\textbf{J1640+2224}) pulsars. 
\textbf{Upper panel}: Violin plots of $\dot{P}_{\rm res}$ using \texttt{Gala2022}. 
MC propagation of asymmetric distance errors revises significance downward ($2.59\sigma \rightarrow 1.93\sigma$), showing that Gaussian assumptions can inflate significance in low-precision systems. We note that $\dot{P}_{\rm GW}$ is negligible and omitted.
\textbf{Lower panel}: Distance PDFs. For \textbf{J1640+2224}, MC sampling (grey dashed) reveals the non-Gaussian long tail from $d=1/\varpi$, while GEP (green) underestimates the uncertainty. }
    \label{fig:distance_comparison}
\end{figure}

We applied this analysis to the six neighboring binary pairs in Table~\ref{tab:residuals_sigma}. 
To rule out local effects like orbital companions or stellar flybys (e.g., the anomaly of J2043+1711~\citep{2025ApJ...983...62D}), robust DM subhalo detection requires simultaneous anomalies in a pair of binary pulsar systems.
None of the pairs show both pulsars with even a $1\sigma$ anomaly, much less the $3\sigma$ significance required for an evidence.
Overall, the observed accelerations are highly consistent with smooth Galactic potential models (see also Fig.~1 in ~\cite{2024PhRvD.110b3026D}), indicating no evidence for a DM subhalo in the current data.

Figure~\ref{fig:distance_comparison} illustrates how distance error treatment affects result credibility, 
comparing two sources from~\cite{2026PhRvL.136d1201C}. 
For the high-precision pulsar \textbf{J1713+0747}, tiny parallax uncertainty ensures robust consistency among three Galactic models, 
with residual deviations of only $0.29\sigma$ to $0.70\sigma$, in agreement with \cite{EPTA2023,Zhu_2018}. 
In contrast, \textbf{J1640+2224} exhibits greater significance. 
While a Gaussian approximation yields a residual significance of $\sim 2.6\sigma$, 
MC sampling of its non-Gaussian distance distribution reduces this to $\sim 2.0\sigma$, which is consistent with \cite{EPTA2023}.
Moreover, its orbital period ($P_b \approx 175$ days~\cite{EPTA2023}) lies close to the semiannual modulation timescale ($\approx 182.6$ days), making orbital parameters possibly to degeneracy with the parallax signature and potentially contributing to the apparent anomaly. 
Based on these considerations, we find no evidence for a DM subhalo signal.

We propose that the candidate signals reported in ~\cite{2026PhRvL.136d1201C} likely arise from two statistical biases.
First, inverting parallax to distance via $d = 1/\varpi$ with symmetric error propagation neglects the long-tailed distance distribution that arises at low precision. 
This systematically underestimates the errors in the Shklovskii acceleration, artificially inflating the significance of the residuals.
Second, their gravitational potential model can be overly simplified. 
The radial force is derived solely from a flat rotation curve, ignoring significant radial and vertical velocity gradients~\citep{Wang_2022,2024ApJ...976..185S}. 
The vertical gravity includes only a stellar disk, omitting contributions from gas and DM, and lacks the radial fall-off in surface density, despite the strong radial dependence of the vertical field~\citep{2024ApJ...960..133G,2024ApJ...976..185S}. 
The spiral arm potential relies on a theoretical model~\citep{Antoja_2011}. 
Although VLBI and Gaia data can map spiral structure, substantial uncertainties in arm geometry persist~\citep{Xu_2023}. 
Incorporating such poorly constrained arms into precision dynamical analyses may introduce additional systematics rather than improve accuracy.
To address these limitations, we adopt data-driven Galactic potentials (see Table~\ref{tab:residuals_sigma}) 
that include all major mass components with observationally motivated radial and vertical profiles.

\section{Summary and Discussion}
We evaluate the detectability of DM subhalos in the solar neighborhood using binary pulsar timing. 
We develop an analytical framework that incorporates the subhalo mass function, LOS acceleration projection effects, and the geometric requirements for joint detection by nearby binaries. 
Applying this framework to the current high-precision binary pulsar sample, we find that the effective detection volumes are strongly suppressed by projection and overlap constraints. 
Even under optimistic assumptions for the local subhalo abundance, 
the expected cumulative detection probability is as low as $P_{\rm det}\lesssim 1.7\times10^{-4}$. 
These results indicate that detecting a $\sim 10^{7}\,M_\odot$ DM subhalo in the solar neighborhood with present pulsar timing data is theoretically highly improbable.

Motivated by this stringent limit, we perform an independent timing residual analysis for 12 high-precision binaries, 
focusing on the six pulsar pairs with separations below $0.5\,\mathrm{kpc}$. 
Our analysis employs a unified MC framework that consistently propagates measurement uncertainties, including the non-Gaussian distance distributions arising from parallax inversion. 
We also evaluate the Galactic acceleration contribution using several widely adopted Milky Way potential models. 
The resulting residual orbital period derivatives are statistically consistent with expectations from smooth Galactic potentials, and none of the binary pairs exhibit correlated acceleration anomalies. 
We find that conventional Gaussian error propagation can significantly overestimate signal significance by neglecting the long-tailed distance distribution,
For PSR J1640+2224, the anomaly significance drops from $\sim2.6\sigma$ using GEP to $\sim2.0\sigma$ under our MC approach.

While the Square Kilometre Array will significantly expand the millisecond pulsar population~\citep{2025OJAp....854256K} , detecting low-mass DM subhalos remains constrained by the slow $\propto T^{1/2}$ improvement in timing sensitivity~\citep{Siemens_2013}. Achieving the $10^{-9}\,{\rm cm/s}^2$ precision necessary to probe subhalos still demands decadal high-cadence monitoring. 
Our simulations demonstrate that even an idealized 100-pair array---with aggressive $0.3\,{\rm kpc}$ separations and $10^{-9}\,{\rm cm\,s^{-2}}$ sensitivity---yields a detection probability below $1\%$, assuming non-overlapping pairwise detection volumes apart from the common Sun-centered component.
Compared to our current 6-pair sample, which largely lacks such precision (see Fig.~\ref{fig:subhalo_joint_detection}), these results suggest that current null detections are a likely consequence of stringent
 geometric and temporal constraints.

In conclusion, our robust null detection and exceptionally low theoretical detection probability establish a critical observational baseline. They resolve the severe theoretical tension introduced by a recently claimed candidate, demonstrating that any genuine detection with current data would require local subhalo abundances drastically exceeding cold dark matter predictions. Furthermore, ruling out this massive subhalo prevents the misdirection of DM indirect detection resources toward spurious targets. The lack of anomalous accelerations also ensures the Galactic phase-space spiral is not falsely attributed to an unverified local perturber. Together, these findings highlight the intrinsic limitations of pulsar-timing subhalo searches and set rigorous constraints for future surveys in the SKA era.

\begin{acknowledgments}

We thank the anonymous referee for constructive comments, and we are grateful to Lei Lei and Qiang Yuan for insightful discussions on pulsar and dark matter physics. 
This work is supported in part by the National Natural Science Foundation of China (No. 12588101), the National Key Research and Development Program of China (No. 2022YFF0503304), the Strategic Priority Research Program of the Chinese Academy of Sciences (No. XDB0550400), and 
the CAS Project for Young Scientists in Basic Research (No. YSBR-092). Y.Z.F thanks the support of New Cornerstone Science Foundation through the XPLORER PRIZE. 
\end{acknowledgments}

\bibliographystyle{aasjournal} 
\bibliography{ref}

\end{document}